# First foot prints of chemistry on the shore of the Island of Superheavy Elements

**R. Eichler**[1,2]

[1]Paul Scherrer Institute, Villigen, 5232, Switzerland
[2]University of Bern, Freiestr.3, 3010 Bern, Switzerland

robert.eichler@psi.ch

**Abstract**. Chemistry has arrived on the shore of the Island of Stability with the first chemical investigation of the superheavy elements Cn, 113, and 114. The results of three experimental series leading to first measured thermodynamic data and qualitatively evaluated chemical properties for these elements are described. An interesting volatile compound class has been observed in the on-line experiments for the elements Bi and Po. Hence, an exciting chemical study of their heavier transactinide homologues, elements 115 and 116 is suggested.

## 1. Introduction

Transactinide elements are produced artificially in heavy-ion induced nuclear fusion reactions. This production path allows for production rates on an atom-at-a-minute to atom-at-a-month scale applying currently available accelerator and target techniques (see for review [1]). The production scheme suffers also from vast amounts of by-products from multi-nucleon transfer reactions and from nuclear fusion reactions with contaminants in the target material and other irradiated construction materials. Hence, a chemical identification of transactinides appears to be only possible if the chemical separation of those by-products is sufficient. The produced transactinides are with some exceptions typically short-lived with half-lives in the sub-millisecond-to-minute time scale. Thus, chemical investigation methods have to be very fast and efficient. The chemical analysis of experiments performed on the one-atom-at-a-time scale requires the knowledge or presumption of defined chemical states unchanged during the chemical procedure. Moreover, the single atoms have to be able to change their chemical state within a chemical equilibrium several times to allow a quantification of thermodynamic data from the equilibrium on the basis of probabilities to find the atom in one or in the other chemical state.

Gas-phase chromatography techniques are nowadays the most efficient and fast methods to investigate chemical properties of transactinides (see for review [2]). Within the methods of gas chromatography, thermochromatography is the most efficient one. Thermochromatography probes the adsorption interaction of a volatile species with a defined stationary phase in a temperature gradient. The species are transported by the mobile carrier gas phase. Numerous encounters of the species with the surface lead to a retention time which is determined via the life time of their radioactive decay for short-lived isotopes and which is equal to the experimental duration for long-lived isotopes. This retention leads to an accumulation of decays or an accumulation of long-lived activity at a certain temperature within the gradient. Using models of gas chromatography [2,3] the interaction enthalpy of the species with the surface can be deduced from this deposition temperature.

The chemical identification of the transactinide species is usually performed indirectly via comparison to the chemical behavior of homologues in the corresponding groups of the periodic table at the same experimental conditions. The use of similarly low concentrations is of outmost importance and it is realized by applying carrier-free amounts of the homologues. Correlations are established between the observed adsorption behavior to macroscopic properties, such as, e.g. sublimation enthalpy or boiling points etc. [3] to bolster the speciation. In some cases density-functional theory is able to calculate adsorption properties of transactinides. Not being "ab-initio", usually these calculations are "optimized" to describe the adsorption behavior of the homologous species (see for review [4]).

Transactinide elements Rf, Db, Sg, Bh, and Hs have been chemically identified as typical members of their corresponding groups 4-8 of the periodic table (see for review [5]). Here, the investigation focused on the most stable chemical compounds at the highest oxidation states. The trends established by the periodic table suggest an increasing stability of the highest oxidation states in these transition metal groups with increasing atomic number. Otherwise, a decreasing volatility of the compounds in the highest oxidation states is predicted due to a stronger stabilization of the solid state compared to the gaseous state [3]. Indeed, the first gas phase chemical investigations of the light transition metal transactinides in the form of $RfCl_4$, $DbCl_5$/$DbOCl_3$, $SgO_2Cl_2$, $BhO_3Cl$, and $HsO_4$ confirmed these trends (see Fig.1).

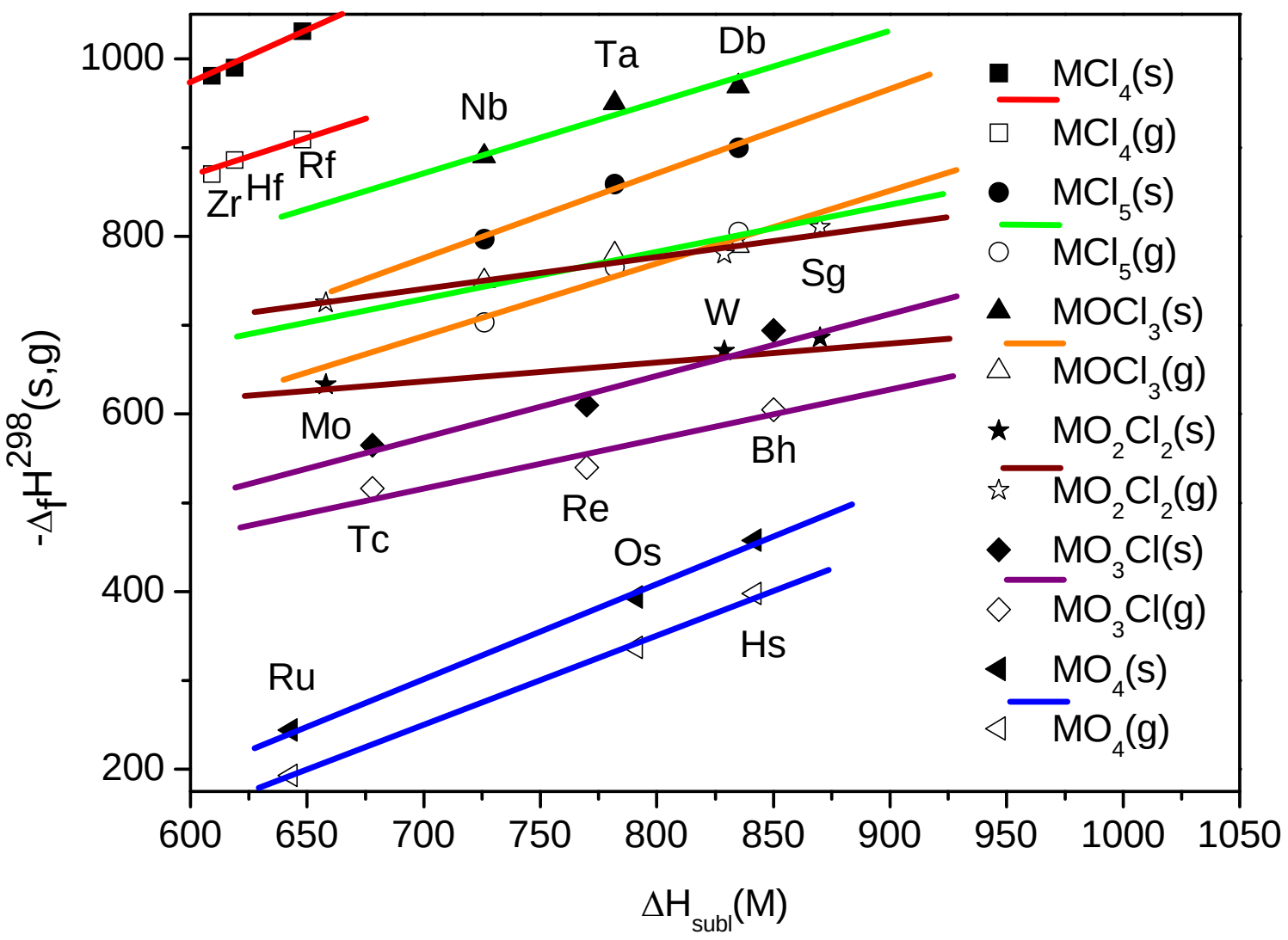

**Fig.1** The standard formation enthalpies of transition metal compounds of the rows 5, 6, and 7 of the periodic table in their maximum oxidation state in the solid state (closed symbols) and gaseous state (open symbols) as a function of the sublimation enthalpy of the corresponding element. The lines are used to visualize the trends. Note here the increasing difference between the upper and the lower line of the same color which represents the sublimation enthalpy of the species representing a measure of volatility.

Since the beginning of this century the reported production of isotopes of even heavier transactinides the superheavy elements with half-lives in the order of seconds (see for review [6]) lead to large excitement also among nuclear chemists. For the first time superheavy elements became assessable to chemistry. In this paper we will summarize the efforts made towards their chemical investigation. The s- and p- electron shells are the valence shells in theses elements responsible for most of the chemical properties. These electron shells are subject to large influences by direct and indirect relativistic effects. Therefore, the spread of data predicted for chemical properties of SHE is much larger as compared to the transactinide transition metals [7-10].

## 2. The chemical investigation of copernicium (element 112)

The chemical investigation of copernicium is of utmost interest to physical chemistry and chemical theory. The valence electrons of this element are situated in the 7s electron orbital. This orbital is subject to strong relativistic contraction. The secondary shielding of the nuclear charge and the spin-orbit splitting lead to a weakening and loosening of the 6d electron orbitals. The question for the chemistry experimentalists was to suggest and to setup a chemical system, which is able to distinguish between a noble metallic and an inert-gas-like copernicium. Already in 1980 it was suggested that the adsorption on noble metallic surfaces shall distinguish between both properties [10].

A thermochromatography detector based on the cryo-online detector technique [11] applied during the experiments with $HsO_4$ [12] was rebuilt and optimized to serve for adsorption studies of SHE on noble metal surfacse [13] (see Figure 2).

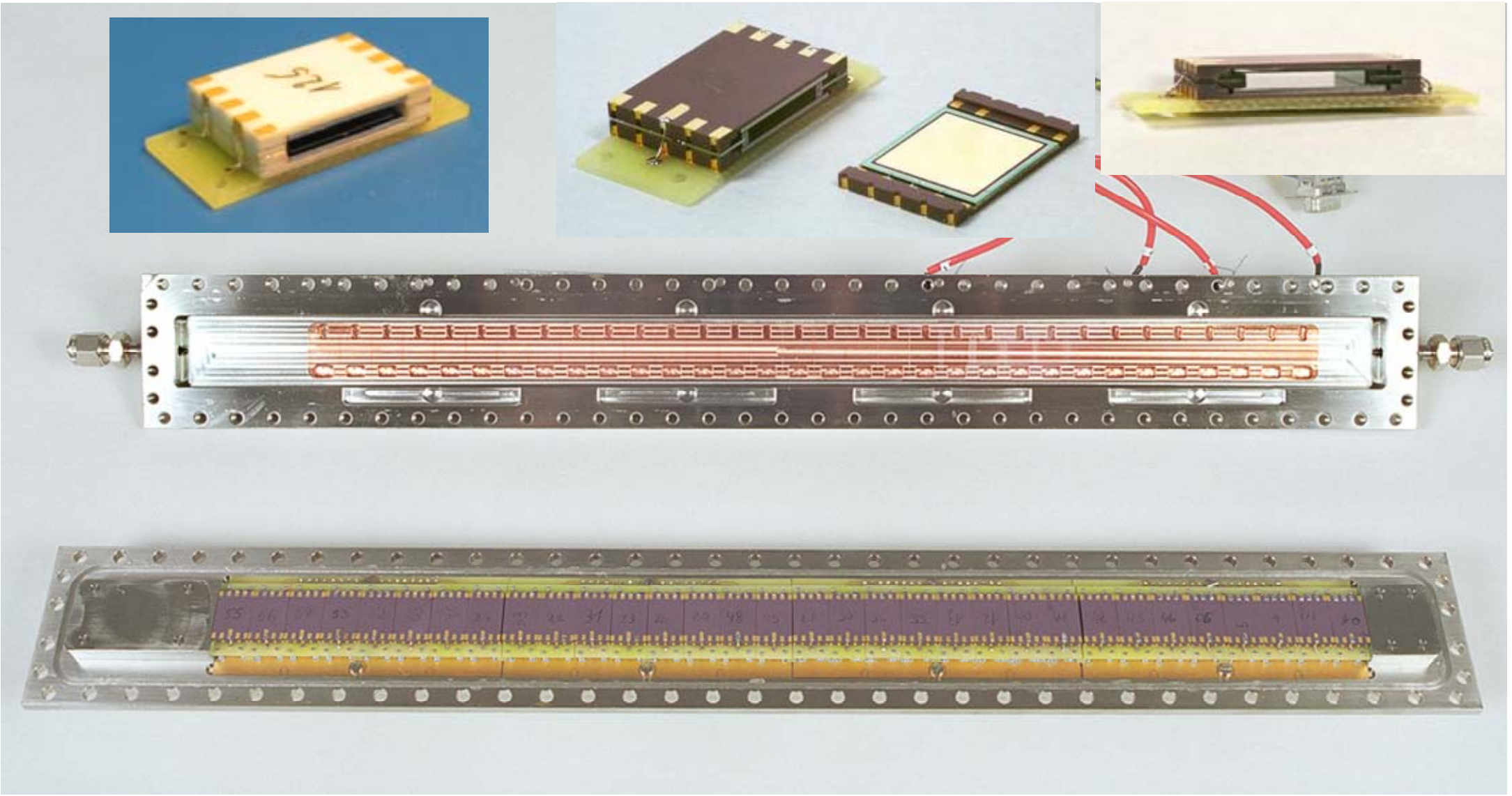

**Fig. 2** The Cryo-Online-Detector COLD. The array of 32 sandwich-type detectors (upper side gold-covered) is shown mounted inside the channel. The lid contains the gas inlet and outlet and a copper bar welded into the steel housing and pressed on the top detectors to provide a stable almost linear temperature gradient of about 5 K/cm. The sandwich-type detectors made at PSI (center and right and mounted in the channel) and at ITE Warsaw (left) are shown as inserts in the upper part of the figure.

Gold was suggested to be the stationary surface of choice for the investigation, because it is assumed to be stable against surface oxidation. Thus, it should provide a clean metallic surface throughout the month-long experiments. Hence, the detector sandwiches in the new PSI Cryo-Online-Detector (COLD) were covered on one side with a gold layer of about 50 nm thickness. Furthermore, a self-drying gas loop system was developed to be able to run the experiments with a temperature gradient down to -180°C. This gas loop system contains drying cartridges based on Sicapent® and hot tantalum based getter ovens to remove water and oxygen in the carrier gas down to sub 100 ppb levels corresponding to dew points below -100°C. The dew point of the carrier gas has been monitored on-line by dew point sensors.

The operation principle of the entire setup is shown in Figure 3. The nuclear reaction products recoil out of the stationary target with the momentum of the beam and are thermalized in the carrier gas (typically a mixture of 70% He and 30% Ar) within the recoil chamber. The products are flushed into the first rough chemical separation stage consisting of a quartz tube heated up to 850°C with a piece of tantalum metal inside and a quartz wool plug for filtering aerosol particles produced in beam induced sputtering processes. In this first stage separation reactive gases and non-volatile elements are retained together with the aerosol particles. Only volatile and inert products enter a 4 m-long PFA® Teflon capillary held at room temperature (or eventually at 70°C), and connected directly to the inlet COLD detector. This transport represents the second stage separation. Here, volatile species are separated according to their adsorption interaction with the Teflon surface. Hence, only elements and species not strongly interacting with Teflon manage to pass to the COLD. The third stage separation occurs in COLD, where the volatile species are separated according to their adsorption interaction with the gold covered detector surface. The detectors are connected to a sophisticated energy resolving alpha and spontaneous fission fragment spectroscopic system operating in an event-by-event mode to be able to identify time correlated genetically linked decay chains form single atoms of transactinides.

The isotope of copernicium with the mass number 283 was chosen [6]. With a half-life of 4 s and its 9.5 MeV alpha decay to the short-lived (120 ms) mainly spontaneously fissioning $^{279}$Ds it seemed to be ideally suited for gas-phase chemical investigations. The production via the immediate alpha decay of the short-lived complete fusion-evaporation product $^{287}$Fl produced in the $^{48}$Ca induced nuclear fusion reaction with $^{242}$Pu promised in 2004 higher production cross section of about 5 pb [14] compared to the direct production in the reaction of $^{48}$Ca with $^{238}$U, where less than 1 pb is expected [15].

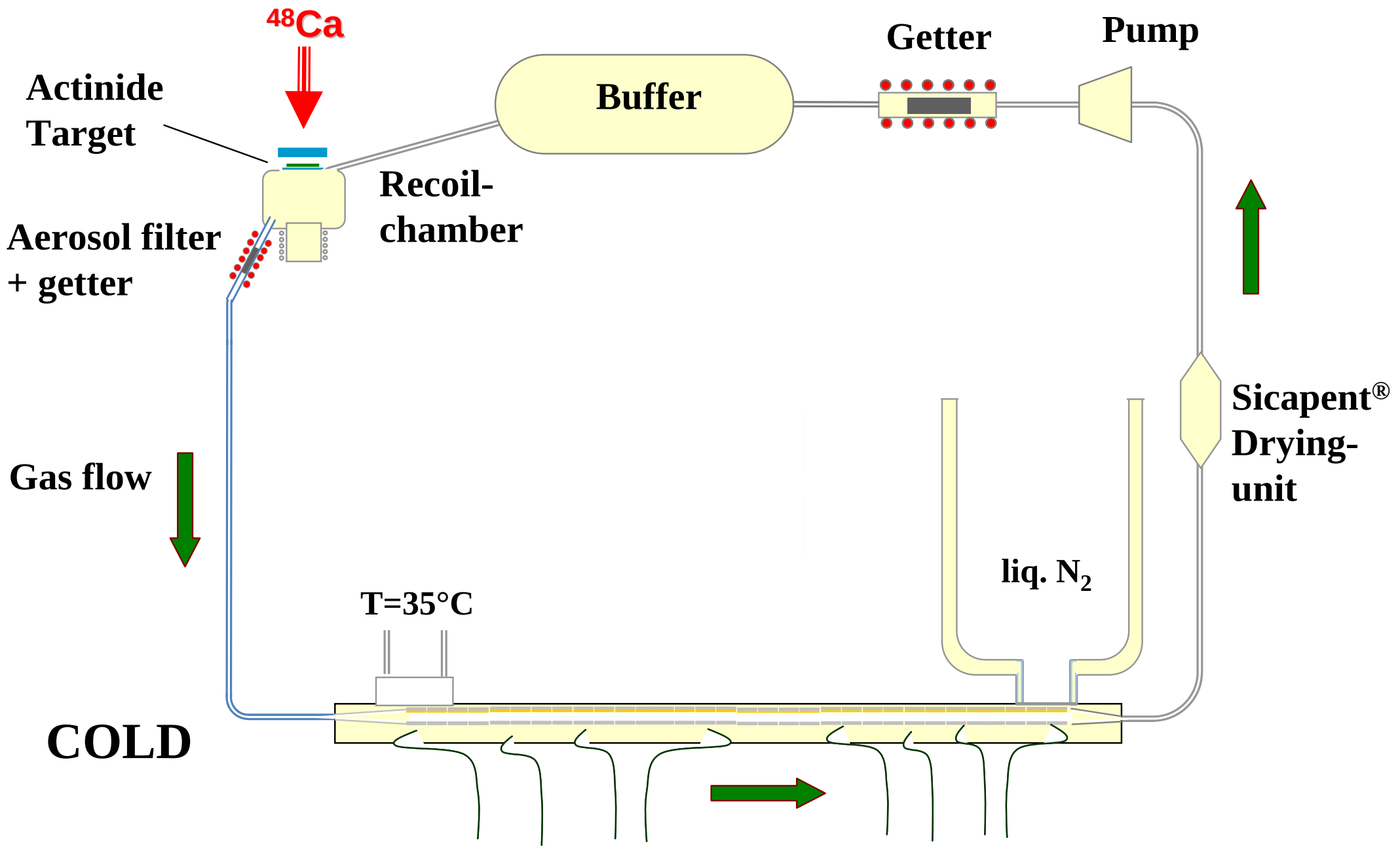

**Fig. 3** Experimental scheme operated for the adsorption thermochromatography with SHE (adopted from [16]).

In 2006 and 2007 two experiment campaigns have been performed using an overall beam dose of 6.2 * $10^{18}$ Ca particles on $^{242}PuO_2$ targets (1.5-2 mg/cm$^2$) prepared on 2 μm Ti-foil backings using the painting technique at the U400 cyclotron at FLNR Dubna, Russia. Altogether, five decay chains related to $^{283}$Cn have been observed (Fig.4, chains 1-5) [16,17].

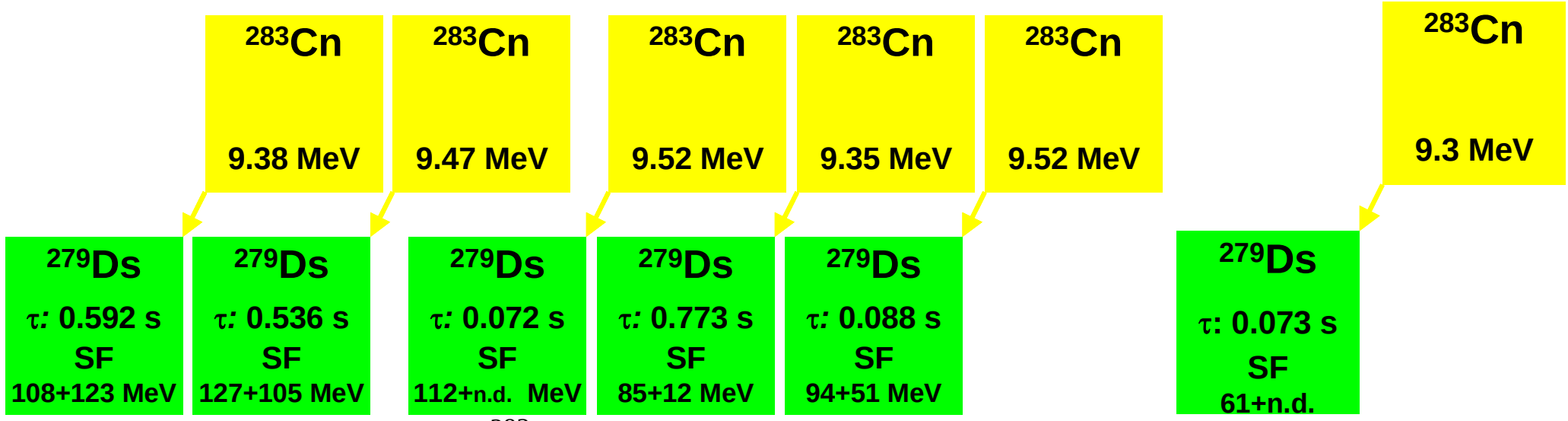

**Fig. 4** Six decay chains related to $^{283}$Cn observed during the chemistry experiments in 2006-2009.

Their deposition pattern along the temperature gradient in the COLD detector is shown in Fig. 5. The observed behavior under three changed experimental conditions show clearly a reversible mobile adsorption process involved in the thermochromatography of elemental Cn on gold. In the first part of the experiment the temperature gradient started at -24°C on the first detector and went down to -180°C. The gas flow was held at about 900 ml/min (Fig.5, upper panel). One decay of $^{283}$Cn was observed at -28°C on detector #2. In the second part of the experiment the temperature gradient was increased by heating the first detector to 35°C. At otherwise equal conditions a second decay of $^{283}$Cn was observed on detector #7 (-5°C) (see Fig.5, middle panel). In the third part of the experiment at the temperature conditions of the second part of the experiment the gas flow was increased up to 1500 ml/min (see Fig.5, lower panel). A broader exponential diffusion-controlled distribution of $^{185}$Hg and almost a non-adsorption of the main part of $^{219}$Rn in the COLD array indicate the higher gas flow. At these conditions three more decay chains of $^{283}$112 were observed on detectors #11 (-21°C, chain 3), 12 (-39°C, chain 5), and 26(-124°C, chain 4) indicating also a sensitivity of the Cn adsorption on the gas flow. An adsorption enthalpy of Cn on gold at zero surface coverage was deduced to as $\Delta H_{ads}^{Au}$(Cn) = -52(+3,-2) kJ/mol using a Monte-Carlo simulation of gas phase chromatography [2], representing the first ever natural constant determined for Cn. This observation was confirmed in two additional experiments aimed at the chemical investigation of Fl (see section 3). There, one additional decay chain from $^{283}$Cn was observed on detector #5 (-7°C) (see Fig.4, chain 6) together with two interesting high energy coincident SF signals on detectors #9 (117/94 MeV, -8°C) and #14(111/101 MeV, -16°C) in the temperature region where Cn was expected [18]. Recently, in an experiment performed at GSI Darmstadt [19,20] $^{285}$Cn revealed an adsorption behavior presumably on a gold surface very similar to these observations. In the view of the results with Fl discussed later, it can not be excluded, that the observation of $^{283}$Cn at -126°C (Fig.4, chain 4) is related to a chemical transport of $^{287}$Fl.

The clear chemical identification of copernicium allowed for application of a correlation established to connect the microscopic atomic property adsorption enthalpy with the macroscopic property sublimation enthalpy [22]. Cn was shown (see Fig. 6) to be a very volatile member of group 12 of the periodic table, with a very weak metallic character [17].

In 2006 the chemical identification of $^{283}$Cn was considered the first independent confirmation of the formation of superheavy elements in the $^{48}$Ca induced nuclear fusion reactions with actinides claimed since 1999 at FLNR, Dubna [23]. Nowadays, these experiments bolster the atomic number of one member of the decay chains from elements Fl and Lv supporting their discoveries at FLNR.

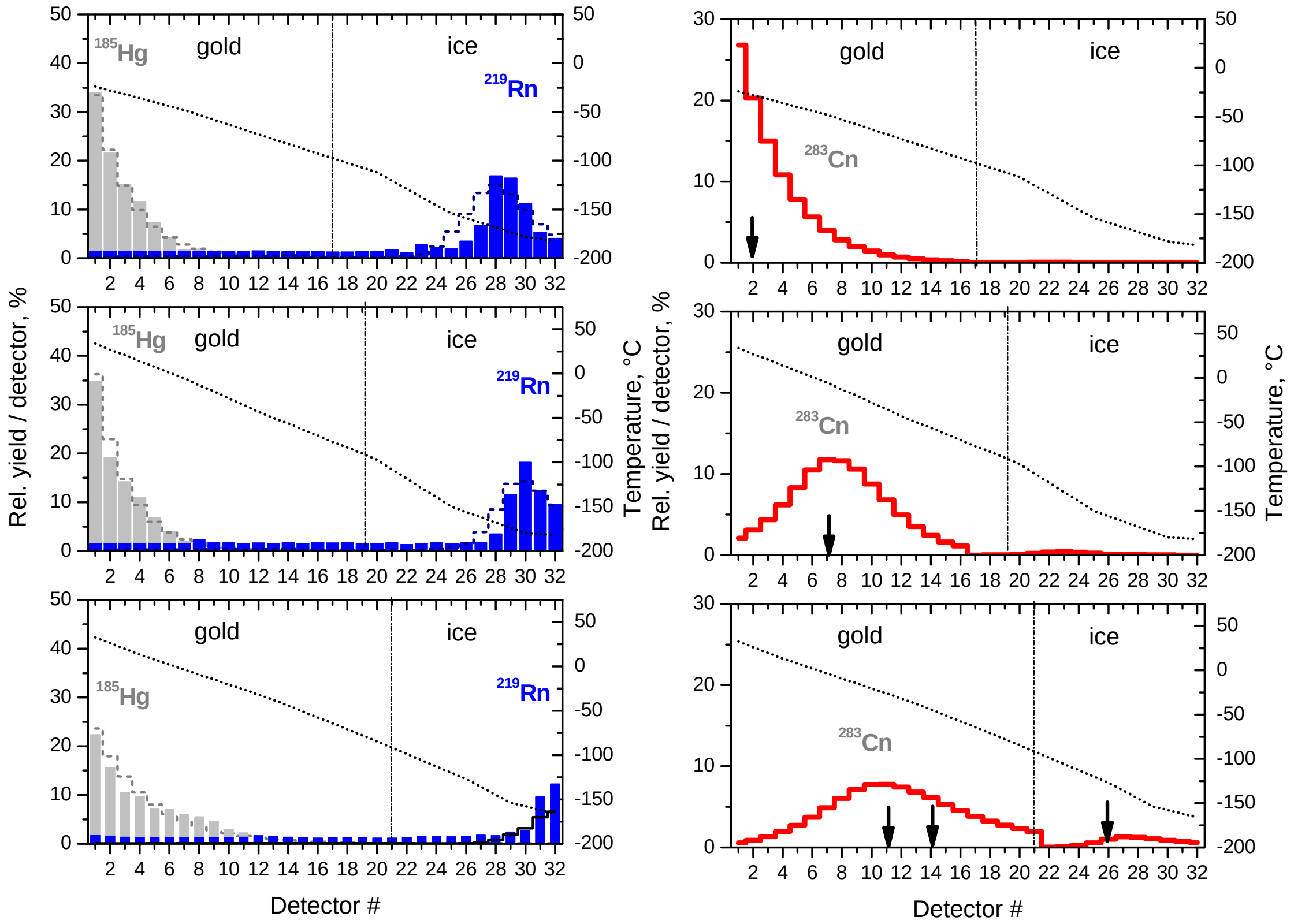

**Fig. 5** The distributions of $^{185}$Hg (grey bars), $^{219}$Rn(blue bars), and $^{283}$Cn(black arrows) at the changed experimental conditions throughout the experiments. The red stepped line indicates the behavior expected by Monte-Carlo simulation of thermochromatography for $^{283}$Cn at these experimental conditions having the adsorption enthalpy on gold as $\Delta H_{ads}^{Au}$(Cn) = -52 kJ/mol. The grey lines are the corresponding simulations for $^{185}$Hg ($\Delta H_{ads}^{Au}$(Hg) =-98 kJ/mol) and $^{219}$Rn ($\Delta H_{ads}^{Au}$(Rn) =-20 kJ/mol).

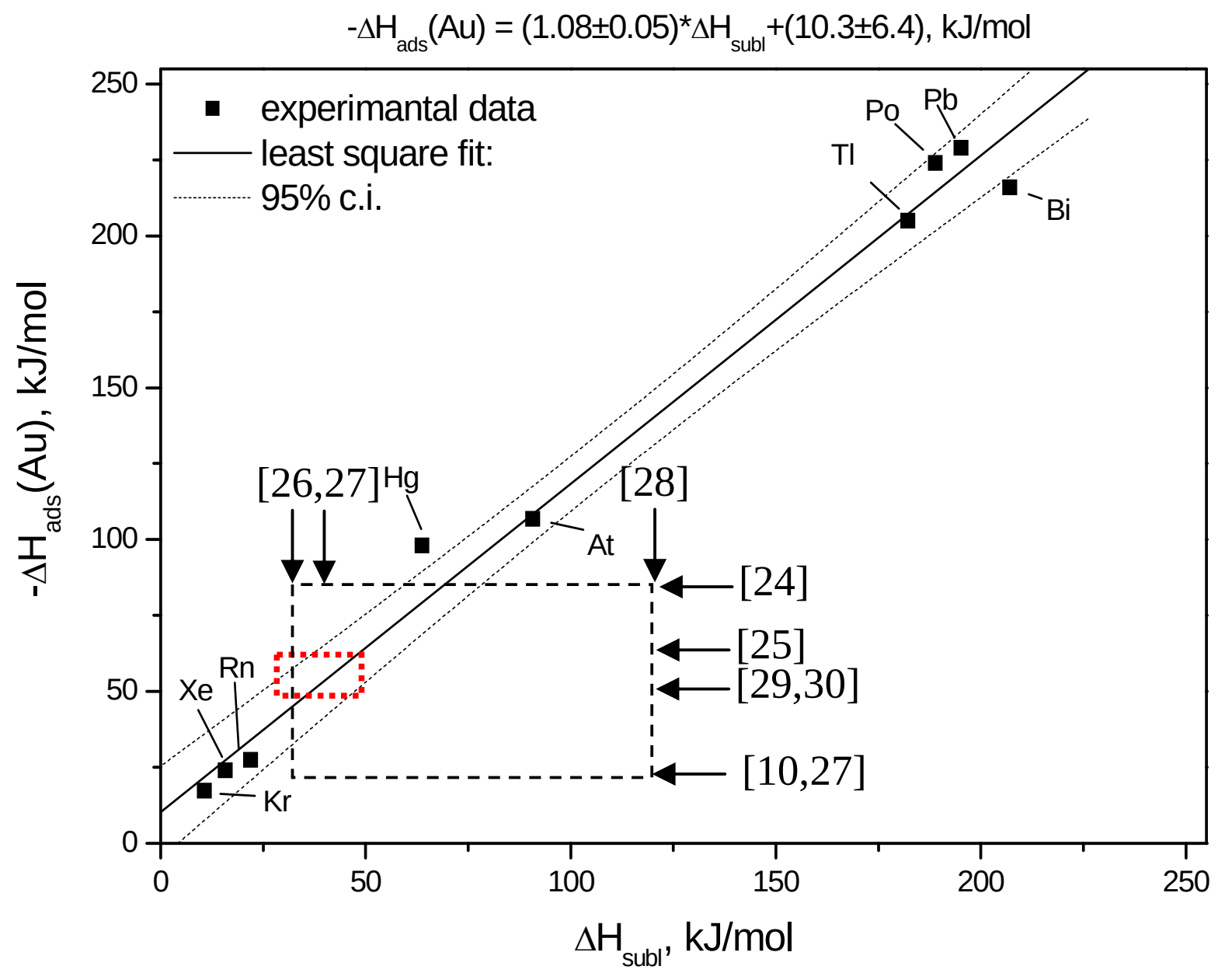

**Fig. 6** Correlation of the microscopic property standard adsorption enthalpy with the macroscopic property related to volatility, the standard sublimation enthalpy adapted from [22]. Macroscopic properties of Cn were deduced empirically from the interaction of Cn with gold (red dotted line) using correlation techniques described in [3,21,22]. Horizontal black arrows show the results of theoretical predictions for the Cn-Au adsorption interaction and vertical black arrows show the predicted standard sublimation enthalpy of Cn, partly disagreeing with the experimental results.

## 3. The chemical investigation of flerovium (element 114)

Flerovium represents the heaviest member of group 14 of the periodic table. Its closest homologue is lead (Pb). From its position in the periodic table Fl is supposed to have fully occupied 7s and $7p^{1/2}$ electron orbitals as chemically relevant valence shells. Both shells are subject to an increasingly strong relativistic contraction. Thus, a chemical inertness close to inert gases was postulated for Fl in 1975 [9]. An increasing elemental volatility can be expected from the trends in the groups 1, 2, 12-16 of the periodic table [26]. However, the adsorption bond properties of Fl on gold surfaces have been predicted by empirical correlations [26,27] and density functional theory [32,33,35] to be metallic.

The experimental setup as used for the chemical identification of copernicium would be suitable to investigate a inert and volatile Fl. In the view of the latest predictions quite unexpectedly, one decay chain (Fig. 7, decay chain 1) was observed simultaneously with the Cn chains described in section 2 on detector #18 held at -88 °C. This decay chain was identical to the signature observed for $^{287}114$ in the discovery experiments of Fl and Lv at the Dubna Gas-filled Recoil separator [6,14].

To bolster this observation the experiment was repeated using the same setup as described in section 2 but using $^{244}Pu$ targets thus aiming at the production of the more long-lived $^{288}Fl$ and $^{289}Fl$. The initial observation was confirmed by the detection of two further Fl atoms (Fig. 7, decay chains 2 and 3) on detectors #3 (-4°C) and #19 (-90°C).

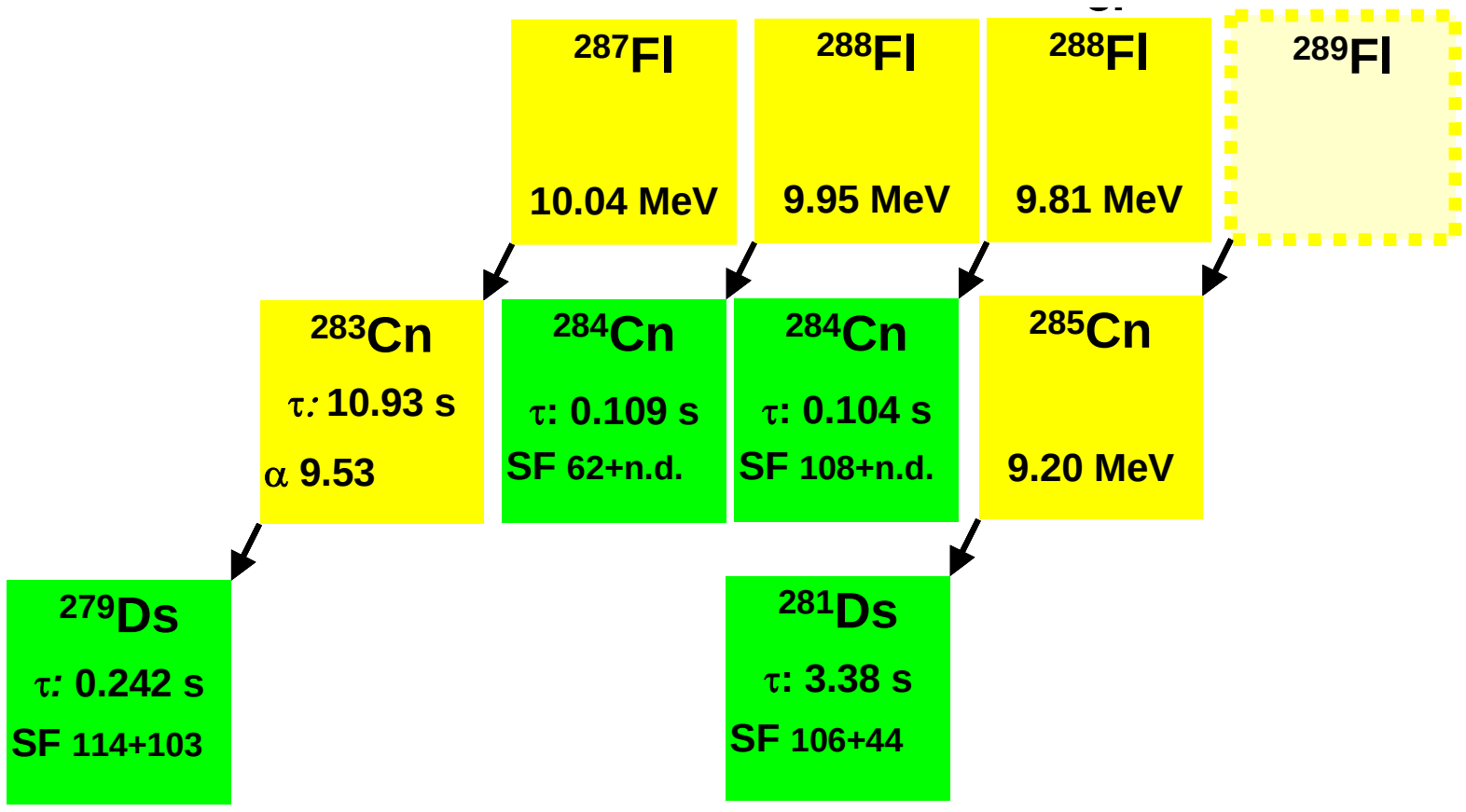

**Fig. 7** Four decay chains presumably related to the chemical observation of Fl adsorbed on gold.

The isotope $^{289}$Fl could not be unambiguously identified in these experiments due to the too high background within the long correlation time required for an efficient detection. For this reason the experiment was repeated using the DGFRS as physical preseparation device [18]. Connecting otherwise unchanged setup, shown in Fig. 3, to the focal plane exit of the DGFRS via a dedicated recoil chamber this preseparation guaranteed highest background reduction. However, a considerable drop in yield was observed. Within a two-month experimental campaign a beam dose $9.72 \cdot 10^{18}$ $^{48}$Ca particles was passed on an about 300 μg/cm$^2$ $^{244}PuO_2$ target electrochemically deposited on a 1.75 μm Ti backing. In this experiment another decay chain was observed (Fig. 7, chain 4) on detector #19 held at -93°C. Even though, the α-decay of $^{289}$Fl was missing (probability is about 20%) in this chain its position in the detector array indicates a confirmation of the previous chemical identification of Fl (see Fig.8, lower panel).

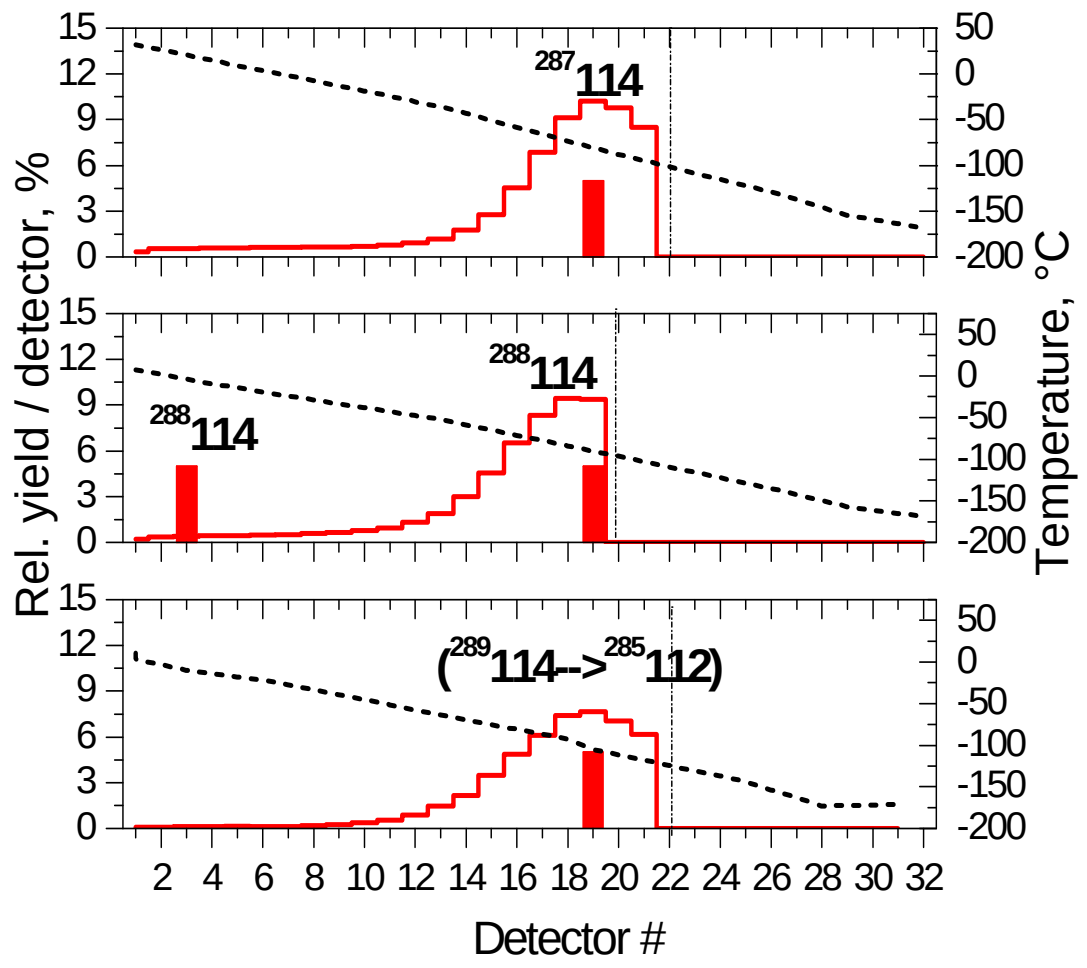

**Fig. 8** Thermochromatograms of three different isotopes of Fl investigated using COLD with one-side gold covered detector surface at similar gas-flow conditions with slightly changed temperature gradients (black dashed lines).The ice-gold border deduced from the dew point in the carrier gas is indicated by the vertical dashed-dotted black line. The red lines represent the Monte-Carlo simulation [2] of the expected deposition of the three Fl isotopes at the respective experimental conditions using the $\Delta H_{ads}^{Au}(Fl) = -38$ kJ/mol.

The deposition pattern observed for all events tentatively attributed to Fl adsorption on gold are shown in Fig.8. The consistent description of the adsorption of all three isotopes of Fl by the Monte-Carlo simulation of thermochromatography [2] using only one adsorption enthalpy value is an indication for a very weak adsorption interaction of element 114 with gold [31], not expected by theory [30,32,33,34]. Figure 9 summarizes the expectations by theory and shows the discrepancies to the experimental work. Recently, this empirical approach was criticized [32]. Therefore, it shall be noted here again, that this correlation is valid only for adsorbed elements on gold with small net-adsorption enthalpies, meaning with no strong chemical interaction with the surface i.e. under preservation of the original chemical state upon adsorption. This is true for the adsorption of the light elements on gold providing the correlation and it can be assumed true for Cn and Fl [27]. Interestingly, the sublimation data [33] and adsorption data on gold predicted for Fl in [32,35] are well consistent with the correlation. Recently, two decay chains attributed to $^{288}$Fl and $^{289}$Fl were attributed to the observation of its adsorption presumably on a gold surface at room temperature in an experiment performed at GSI Darmstadt [19,20]. The conclusion from all these experiments is clearly, that only first steps are made towards the extremely interesting chemical identification of Fl and that more experimental data is required to resolve the Fl-Au interaction behavior. The experiments revealed also, that target stability is one of the main issues to be solved in the future (see section 6).

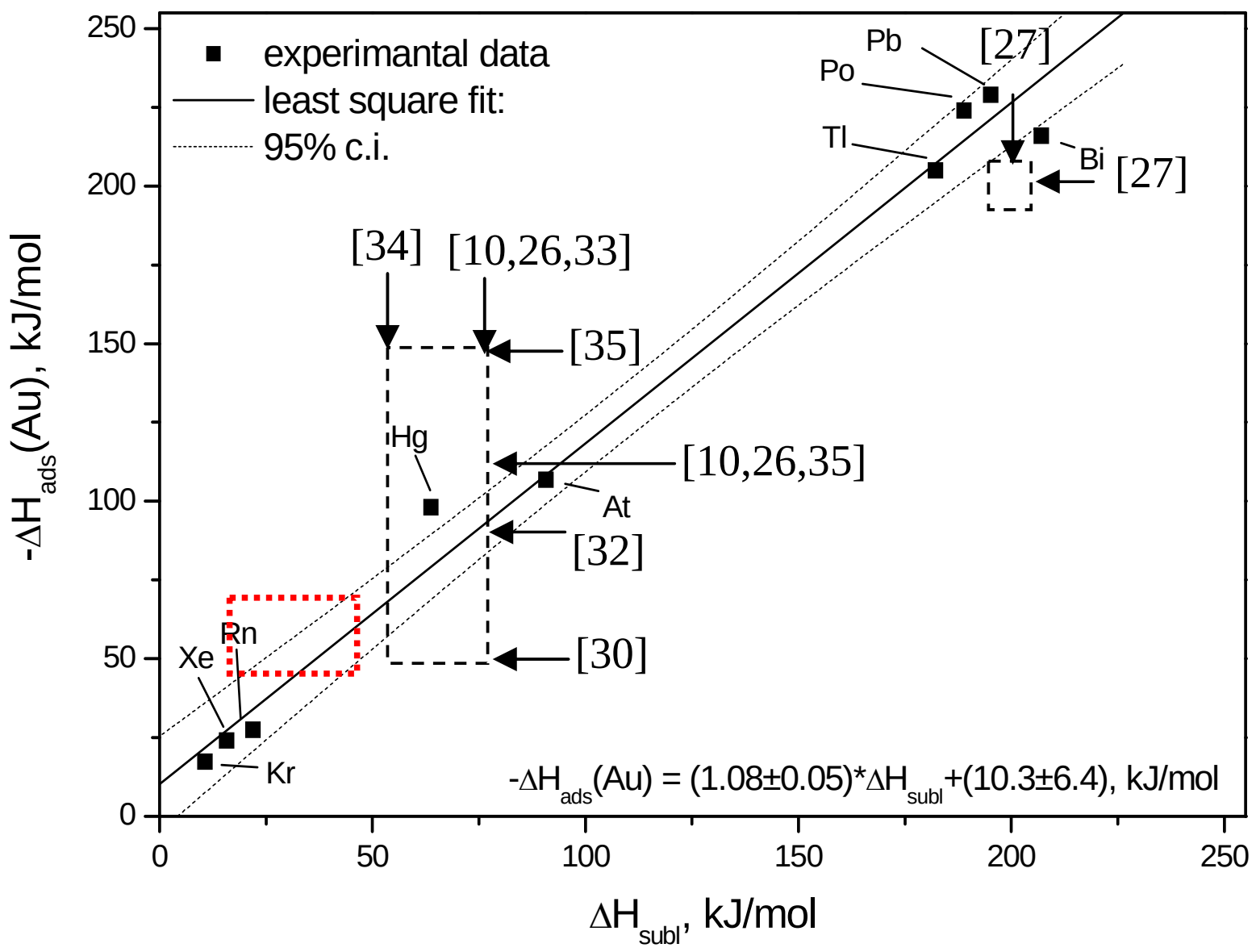

**Fig. 9** Correlation of the microscopic property standard adsorption enthalpy with the macroscopic property related to volatility, the standard sublimation enthalpy adapted from [22]. Macroscopic properties of Fl deduced empirically from the experimentally deduced adsorption interaction of Fl with gold (red dotted square) using correlation techniques described in [3,21,22]. Horizontal black arrows show the results of theoretical predictions for the Fl-Au adsorption interaction and vertical black arrows show the predicted standard sublimation enthalpy of Fl.

## 4. The attempts to chemically investigate element 113

The investigation of element 113 arose from the observation of three isotopes of element 113 having half-lives equal or longer than 0.5 s [36]. In 2002 the isotope $^{284}113$ ($T_{1/2}$=0.5 s) was observed in the decay chain of $^{288}115$ produced in the nuclear fusion reaction of $^{48}Ca$ with $^{243}Am$. Meanwhile, the maximum cross section of this reaction has been measured as 8.5 pb [37]. In 2010 the discovery of element 117 [38] revealed in the observed decay chains of $^{293}117$ and $^{294}117$ two isotopes of element 113 $^{285}113$ and $^{286}113$ with half-lives of 6 s and 21 s, respectively. The predictions for the volatility of elemental E113 show a large spread [10,26,27,33,39]. Again, very interesting is the consistency of the relativistic calculations of the sublimation data (150 kJ/mol) [33] and adsorption data on gold (-159 kJ/mol) predicted for E113 in [40] with the heavily discussed correlation obtained in [22] (see e.g. Fig.9).

The experimental techniques presented here for the investigations of Cn and Fl can be used to confirm or to exclude the higher volatility range of element 113. Thus in 2010 and 2011-2012 four experiment series were performed. Three experiments were performed using the Dubna gas-loop system with isothermal gold covered detectors [41] and with Teflon transport capillaries held at 70°C connecting the recoil chamber and the detectors. One of the experiments was performed using the COLD setup as shown in Fig.3 but also with a heated capillary (70°C). Two experiments performed with the FLNR setup and the experiment with COLD employed the nuclear fusion reaction of $^{48}Ca$ with $^{243}Am$. Altogether, beam doses of $3.2*10^{18}$ (isothermal detector at 0°C), $4.7*10^{18}$ (two isothermal detectors in a row held at 20°C and at 0°C, respectively) and $5.6*10^{18}$ (COLD, no getter oven in the loop and no Ta metal in the filter oven at the recoil chamber, and a temperature gradient established from 35°C down to -110°C) have been applied to initially 800-1500 μg/cm$^2$ $^{243}AmO_2$ targets prepared on 2 μm Ti-foils by the painting technique. The fourth experiment was performed directly after the first experiment using the isothermal detector held at 0°C. Here, about 600-800μg/cm$^2$ $^{249}Bk_2O_3$ targets have been prepared by molecular plating on to 2 μm thin Ti-foils.

Considering reaction cross-sections [37] and all known efficiencies, e.g., target-grid transmission, transport velocity, detection efficiency, these experiments missed to observe roughly 10-20 $^{284}113$ events, which is indicative for either a very low volatility of E113 similar to noble gases or more likely to a lower volatility of the elemental state of element 113, not allowing for an efficient transport of E113 through Teflon capillaries at 70°C. This estimation suffers from uncertainties connected to the observed target damage described in section 6. In the experiments with the $^{249}Bk$ target at an applied beam dose of $9*10^{18}$ $^{48}Ca$ one decay chain was observed [41], which is quite similar to the decay chain of $^{286}113$ observed in the decay pattern from $^{294}117$ [38], representing the 3-n evaporation channel. Unfortunately, the U-400 cyclotron was limited in $^{48}Ca$ beam energy leading to excitation energies between 27 and 35 MeV in the Bk targets only. Therefore, no 4-n evaporation channel could be observed, which would have an estimated factor of four higher cross section at 39 MeV excitation energy [38].

One tentative explanation for all these observations for element 113 is a formation of E113OH needed for a transport of element 113 to the detection devices through the Teflon capillary at 70°C. This formation might be kinetically hindered leading to the observation of its formation only for the longer-lived isotopes of element 113 ($^{286}113$). Also here, further work is required to bolster this conclusion. One way to facilitate the chemical reaction would be adding more water into the carrier gas of the gas phase chromatography setups. One shall be aware, that this approach is limited by the use of Ti-foils as vacuum window and target backing. An increase of $^{48}Ca$ energy form the U-400 cyclotron is expected to be available until end of 2012 [41]. This could allow for detecting the more abundantly produced $^{285}113$ in chemistry experiments, though with a shorter half-life of 6 s. However, to investigate the elemental state of E113 in the gas phase most probably the vacuum chromatography approach [42] is more suitable.

## 5. The observation of a volatile compound class for Bi and Po as models for elements 115 and Livermorium (Lv, element 116)

In 2011 two experiments were performed within two month with Fl and E113 using the COLD setup. There were only two marginal differences between the experiments: 1. The targets were either $^{244}PuO_2$ or $^{243}AmO_2$, respectively; 2. The Ta getters were not used in the loop and in the hot aerosol filter at the outlet of the recoil chamber (see Fig.3) in the experiment with E113, to work at an eventually higher water content (dew point -60°C) in the carrier gas to facilitate the formation of the volatile hydroxide species E113OH. In the experiment with Fl the Ta getters were included in the loop and in the aerosol filter oven.

In figure 10 the sum spectra of both experiments are shown. Surprisingly, they looked quite different. A significantly higher and different activity attributed to an obvious transport of Po and Bi isotopes is observed in the $^{244}$Pu experiment considering almost the same beam doses applied with $5.7*10^{18}$ on $^{243}$Am and $4.6*10^{18}$ on $^{244}$Pu. One production paths of the isotopes in the ground states is the decay of heavier isotopes in the Ra-actinide region that are produced in multi-nucleon transfer reactions of $^{48}$Ca with the target material [43]. On the other hand contamination of the target with macroscopic amounts (μg-range) Pb and Bi is easily possible during the target production procedure. Therefore, a direct formation of $^{212m2}$Bi/ $^{212m2}$Po, $^{212m}$Bi/$^{212m}$Po and $^{212g}$Bi/$^{212g}$Po and $^{213}$Bi/$^{213}$Po in nucleon transfer reactions is very likely to be the main production path for all of these isotopes [44] in both experiments.

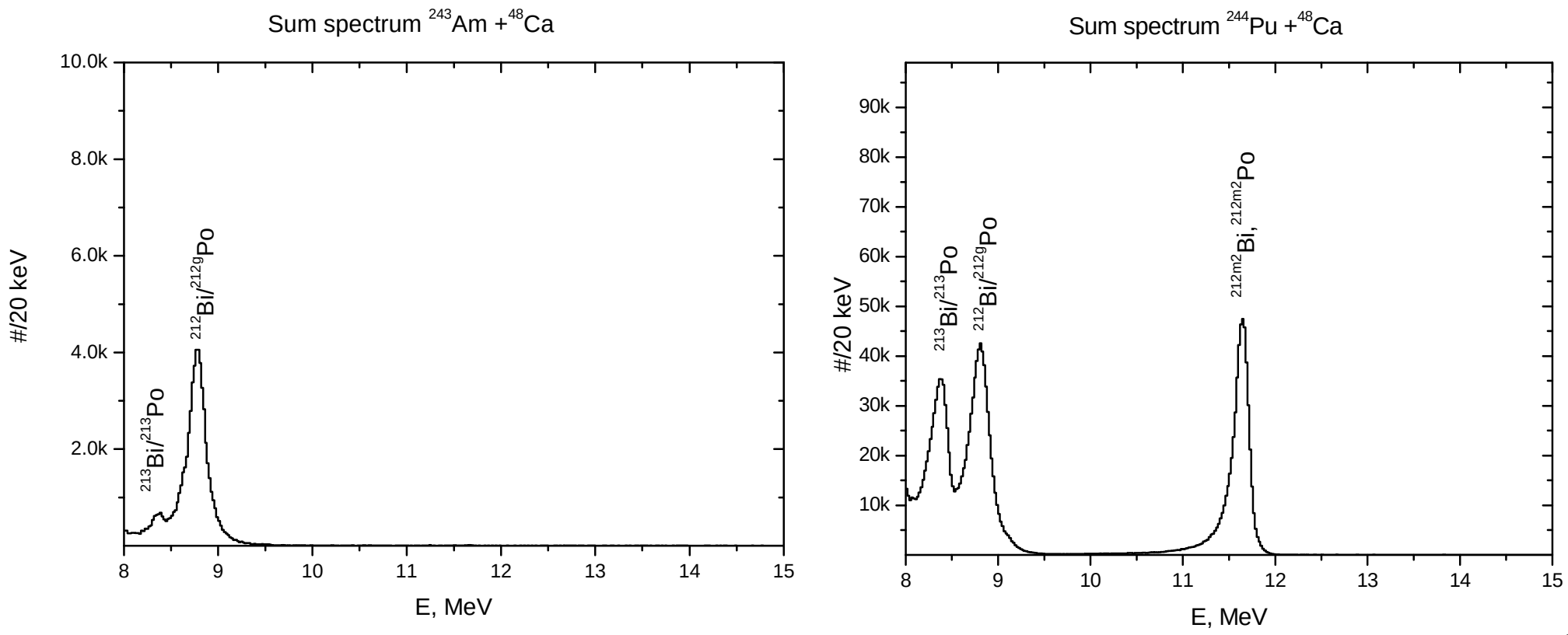

**Fig. 10** Sum alpha spectra above 8 MeV measured throughout the entire experiments with the $^{243}$Am target (left panel) and with the $^{244}$Pu target (right panel).

The only significant difference aside from the target material and its contamination is the use of Ta getters in the $^{244}$Pu experiment. This getter was switched on and off for several times (see Fig.12, both panels, blue line) to check the dependency of the $^{213}$Bi/$^{213}$Po and $^{212m2}$Bi/$^{212m2}$Po transport. Both activities disappear when the getter is switched of. Note here about four orders of magnitude on the $^{212m2}$Po scale (Fig. 11, left panel, red line) and two orders of magnitude the $^{213}$Bi scale (Fig. 11, right panel, green line). Another correlation is related to the dew point measured on-line in the carrier gas (Fig. 11, both panels, black line). If the dew point is below -95°C no efficient transport is observed for both, $^{212m2}$Po and $^{213}$Bi, otherwise the transport efficiency is clearly dependent on the dew point. Thus we conclude that largest transport efficiencies for both elements are observed at high water contents in the carrier gas and with the getter oven switched on. A high water content together with a hot Ta getter in the system leads to a considerable amount of hydrogen in the carrier gas max. ~100 ppm. From these observations, we conclude, that the observed transport of $^{212m2}$Po and $^{213}$Bi is related to the hydrogen content. Already in 1918 Paneth observed the formation of $BiH_3$ with atomic hydrogen in-statu nascendi [45]. The beam induced atomization of hydrogen may yield a considerable amount of atomic H in the recoil chamber which is reacting with the Bi and Po ions recoiling from the target. That the trace amounts of atomic hydrogen yield an efficient transport of Bi and Po points towards an efficient chemical reaction. The question about the reaction velocity remains open for further studies with more short-lived isotopes of Bi and Po.

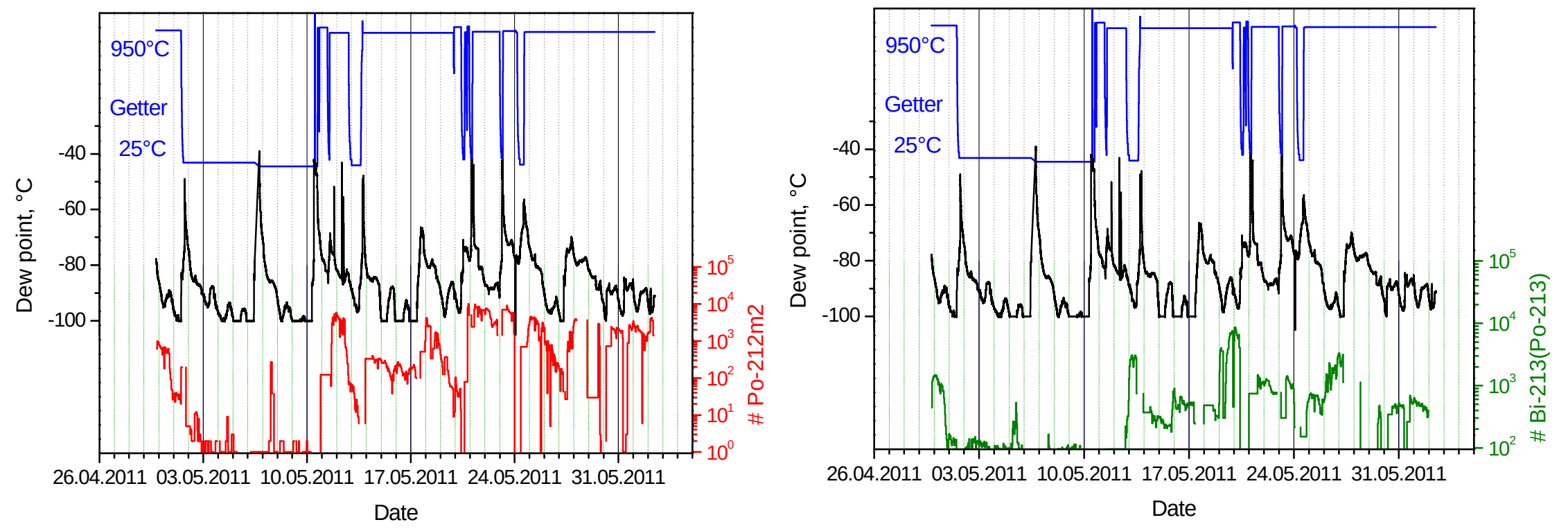

**Fig. 11** Correlation of transported $^{212m2}Po$ (left panel, red line, right hand axis) and $^{213}Bi$ (right panel, green line, right hand axis) with the operation of the loop Ta-getter (blue line, getter temperature on the left side) and with the dew point in the carrier gas (black line left hand axis). Note that the dew point is related to water contents of about 0.01 ppm at -100°C and of about 80 ppm at -40°C, which corresponds to about four orders of magnitude of measured change.

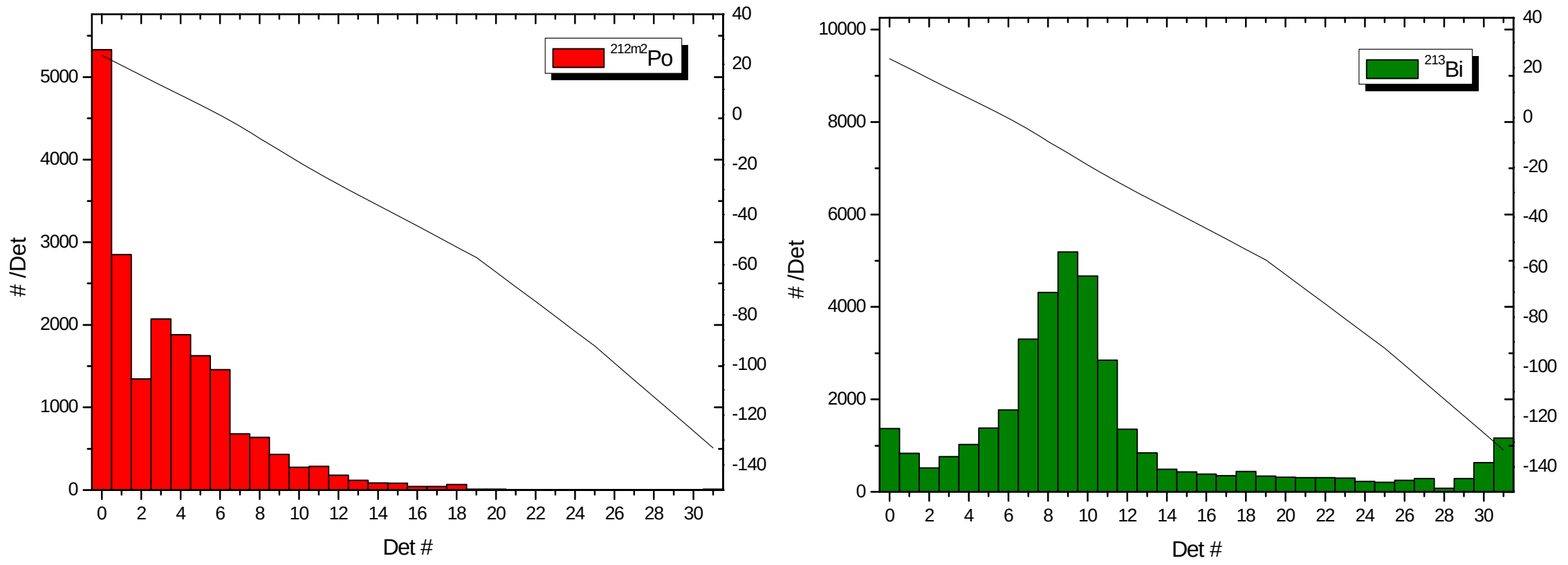

**Fig. 12** Thermochromatograms observed for $^{212m2}PoH_2$ (left panel) and $^{213}BiH_3$ (right panel). The irregularity of the $^{212m2}Po$ activity on detectors #3-6 is related to the transport of $^{212m2}BiH_3$ depositing on the gold in this temperature interval. The increase of $^{213}Bi$ at the end of the detector array is due to the beginning adsorption retention of $^{221}Rn$.

In Fig. 12 the deposition patterns measured for $^{212m2}PoH_2$ and $^{213}BiH_3$ are shown. The deposition temperature dependence on the half-life indicated by the different adsorption position of $^{212m2}Bi$ ($T_{1/2}$ = 9 min) given by the peak on detector 3 in the $^{212m2}Po$ distribution and $^{213}Bi$ ($T_{1/2}$ = 45.6 min) peaking on detector 9 is evidence for a mobile adsorption process of $BiH_3$ on gold. Using Monte-Carlo simulation procedures of thermochromatography based on [46] an limit adsorption enthalpy was estimated for $PoH_2$ on gold as $-\Delta H_{ads}^{Au}(PoH_2)$ >70. For $BiH_3$ an adsorption enthalpy could be evaluated as $-\Delta H_{ads}^{Au}(BiH_3)$ = 65±3 kJ/mol.

The production of Pb isotopes as precursors for a good part of the listed isotopes is likely too. Therefore, the half-lives of the deposited species were roughly determined, after switching off the beam to clearly identify the deposited species. A fast decay of the $^{212}Bi$($T_{1/2}$ = 60.6 min) on the detectors #6-12, where $^{212}BiH_3$ adsorption was observed, excluded the transport via a corresponding $^{212}Pb$ ($T_{1/2}$ = 10.64 h) compound. The fast decay of the main part of $^{212m2}Po$ ($T_{1/2}$ = 45 s) on the first 3 detectors within several minutes excludes its main transport as $^{212m2}Bi$ ($T_{1/2}$=9 min), whereas the part of $^{212m2}Po$ measured on detectors #3-6 decayed with a half-live of about 10 min pointing towards a $^{212m2}Bi$ deposition. $^{213}Po$($T_{1/2}$ = 4.2 μs) is too short-lived to be transported efficiently. The half-life of the deposition attributed to $^{213}Bi$ (see Fig. 12) and measured by the $^{213}Po$ daughter alpha-decay was with ~45 min consistent to the half-live of $^{213}Bi$.
The observation that the transport of the hydride species was possible through quartz wool filter unit held at 850°C with tantalum is very interesting from the stability point of view. Removal of the Ta from this filter unit had no influence on the transport efficiency of the hydride species to the COLD detector. These observations point towards an unexpectedly high thermal stability of the volatile species formed. However, a simple extrapolation scheme along the groups of the periodic table predicts that the hydrides of the element 115 and Lv shall be less stable compared to their lighter homologues. Apart from the very interesting density functional electronic structure calculations for group 16 dihydrides including $LvH_2$ [47] further theoretical calculations are needed regarding the thermodynamic stabilities of $PoH_2$ and $LvH_2$ and regarding the electronic structures and thermodynamic stabilities of $BiH_3$ and $115H_3$.

Otherwise, the predicted reasonable elemental volatility of the SHE 115 and Lv elements and their homologues [26,27,33,40] shall allow also for their investigation using a high temperature vacuum chromatography approach [42] in the future.

## 6. Improvements of experimental techniques

Nowadays, the main limitation for the production of SHE is target stability. Intense heavy-ion beams are delivered by large accelerators. Stationary or rotating targets are irradiated, typically consisting of stable oxide compounds or metals deposited onto thin foils of titanium or carbon. Vapour deposition of metals is possible only if enough target material is available, e.g. uranium or lead. Amounts from single milligrams to several tens milligrams are available for the rare and expensive heavy actinides. Molecular plating of the target material onto the metallic target backing is applied [48]. In this case an oxide layer is deposited. In some cases a painting and baking procedure is applied to obtain oxidic targets on titanium backings. During the intense irradiations of up to 1-4 pμA equivalent to $10^{13}$-$10^{14}$ ions/cm$^2$s the thermal, mechanical, and chemical stability of these targets is crucial. Unfortunately, oxide targets have some disadvantages due to their thermal and electrical insulation properties and due to possible high temperature reactions of the oxide material with the target backing. In figure 13 a photomicrograph of a stationary $^{244}PuO_2$-target (0.6 mg/cm$^2$) electroplated onto a 2 μm thin foil of titanium is shown after irradiation with an integral beam dose of $3*10^{18}$ at intensities of 0.5 pμA. Clearly the chemical and thermomechanical degradation of the titanium backing is visible. The non-irradiated part behind the honey-comb shaped cooling grid indicates the material condition before irradiation.

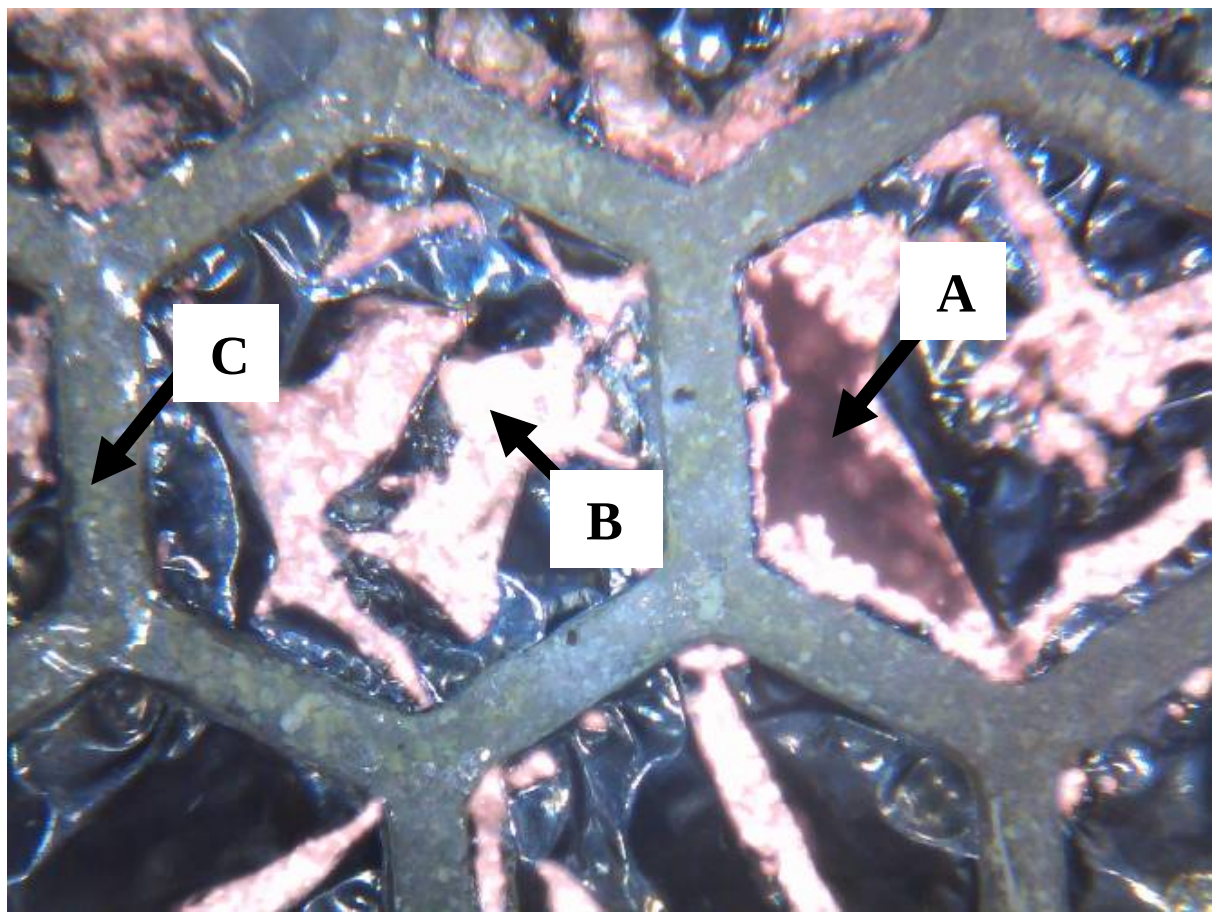

**Fig. 13** Microphotograph of a part of a severely damaged $^{244}PuO_2$ target after intense irradiations (illuminated from the back side) [51]. A: Macroscopic hole in the target material layer and in the backing; B: Target material is still intact, the target backing material is burned, presumably due to the reaction with the target material. C: The target as it looked before irradiation. This part is shielded by a honey comb shaped copper cooling grid during the irradiation.

For the further investigations of superheavy elements using the COLD thermochromatography technique the stability of the heavy ion irradiation targets has to be improved. Recently we proposed new target materials to be used as stationary targets based on intermetallic compounds of actinides with high melting noble metals [49]. The actinide oxide target material is electroplated on thin noble metal foils, such as Pd or Rh. Subsequently, this target is heated up in a pure hydrogen atmosphere to temperatures of about 1000°C. In a coupled reduction process [50] a thin intermetallic target of homogeneous area thickness is formed (see Fig. 14). The intermetallic state has the advantage of heat and electrical conductance and a high chemical stability.

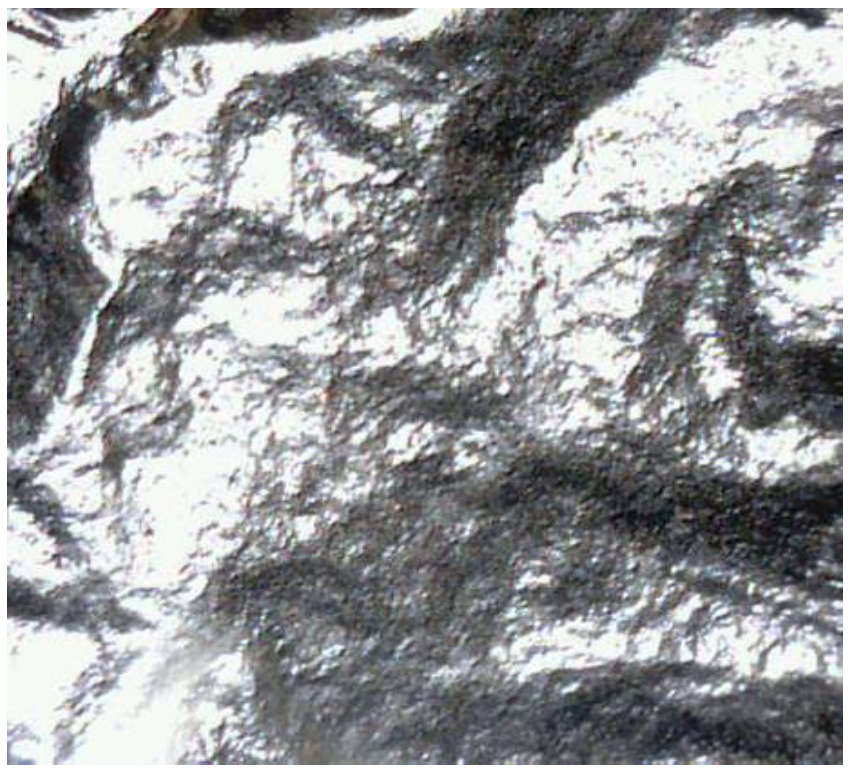

**Fig. 14** Microphotograph of an intermetallic $^{243}$Am (2 mg/cm$^2$)-Pd (3 μm thickness) target before irradiations [51].

A heavy ion irradiation of such an Am$^{243}$/Pd intermetallic target with intense beams of $^{48}$Ca has been performed at FLNR Dubna. First results revealed a considerably higher stability of this target compared to the conventional oxide targets on titanium backings. A detailed analysis is ongoing and will be published elsewhere [51].

## 7. Conclusion

Chemistry has arrived on the Island of Stability of Superheavy Elements. Interesting tasks regarding the confirmation of the very volatile Fl and regarding other chemical properties of Cn and Fl are to be tackled in the future. The investigation of element 113 needs either an increase of water vapor pressure in the carrier gas to facilitate the formation of 113OH expected to be more volatile compared to the elemental state. A new class of stationary targets has been developed for a stable production scheme of SHE, mandatory for their further chemical investigation. The observation of the formation of a volatile hydride species of group 15 and group 16 elements homologous to element 115 and Lv, presumably $MH_3$ and $MH_2$, respectively, opens up an interesting chemical system for their gas-phase chemical investigation. Group 14 element lead was shown not to form a volatile lead hydride compound at conditions where the group 15 and 16 hydride compounds have been observed. More model studies are required to assess the most efficient way to produce these species, though.

**Acknowledgements**

We thank the staff of the U-400 cyclotron for providing intense beams of $^{48}$Ca. The $^{242}$Pu target material was provided by RFNC-VNIIEF, Sarov, Russia. The $^{244}$Pu and $^{249}$Bk target material was provided by the U.S. DOE through ORNL, Oak Ridge, USA. The LLNL work was performed under the auspices of the U.S. Department of Energy by Lawrence Livermore National Laboratory. This work was supported in part by the Russian Foundation for Basic Research and by the Swiss National Science Foundation.